\documentclass[letterpaper, 10 pt, conference]{ieeeconf}

\IEEEoverridecommandlockouts
\usepackage{epsfig}

\title{\LARGE \bf

Dual-sided Peltier Elements for Rapid Thermal Feedback in Wearables
}

\author{Seongjun Kang$^{1}$, Gwangbin Kim$^{1}$, Seokhyun Hwang$^{1}$, Jeongju Park$^{1}$, Ahmed Elsharkawy$^{1}$, and SeungJun Kim$^{1*}$
\thanks{$^{1}$Seongjun Kang, Gwangbin Kim, Seokhyun Hwang, Jeongju Park, Ahmed Elsharkawy and SeungJun Kim are with the School of Integrated Technology,
Gwangjun Institute of Science and Technology, Gwangju, South
Korea; *SeungJun Kim is the corresponding author.}
}

\begin{document}

\maketitle
\thispagestyle{empty}
\pagestyle{empty}

\begin{abstract}
This paper introduces a motor-driven Peltier device designed to deliver immediate thermal sensations within extended reality (XR) environments. The system incorporates eight motor-driven Peltier elements, facilitating swift transitions between warm and cool sensations by rotating preheated or cooled elements to opposite sides. A multi-layer structure, comprising aluminum and silicone layers, ensures user comfort and safety while maintaining optimal temperatures for thermal stimuli. Time-temperature characteristic analysis demonstrates the system's ability to provide warm and cool sensations efficiently, with a dual-sided lifetime of up to 206 seconds at a 2V input. Our system design is adaptable to various body parts and can be synchronized with corresponding visual stimuli to enhance the immersive sensation of virtual object interaction and information delivery.
\end{abstract}

\section{INTRODUCTION}
Interactive technologies have enhanced the XR experience by incorporating multimodal sensations, including olfaction \cite{lee2022auditory, 10.1007/s10055-024-00997-y}, touch \cite{melo2020multisensory, kang2023giant}, wind \cite{kang2023giant2}, and vibration \cite{hwang2023enhancing, hwang2023electrical}. Among these, tactile feedback has received the most attention, aiming to simulate the physical sensation of interacting with virtual objects. Since the sense of touch arises from somatosensory integration, encompassing temperature and pressure, comparable attention should be given to the design of thermal feedback for a more immersive and realistic XR experience. Particularly, the fluctuations in warmth or coolness experienced during dynamic interactions with virtual objects enhance immersion and presence, making them valuable for interpersonal interactions.

Peltier elements are preferred for thermal feedback in XR because of their capability to electrically modulate temperature. While traditional Peltier systems, controlled by current, effectively conveyed the overall temperature \cite{peiris2017thermovr} or atmosphere of a scene \cite{maeda2019thermodule}, they were less effective for swift temperature transitions due to their slow thermal response rate. To tackle this issue, we introduce motor-driven, preheated Peltier elements designed to provide rapid alterations between warm and cool sensations using both sides of the elements. In this study, we present our multi-layered structure, which enables the incorporation of both sides of the Peltier elements to deliver warm and cool  sensations in XR, along with their time-temperature characteristics and potential applications in XR scenarios.

 \begin{figure}[thb]
      \centering
      \includegraphics[width=\linewidth]{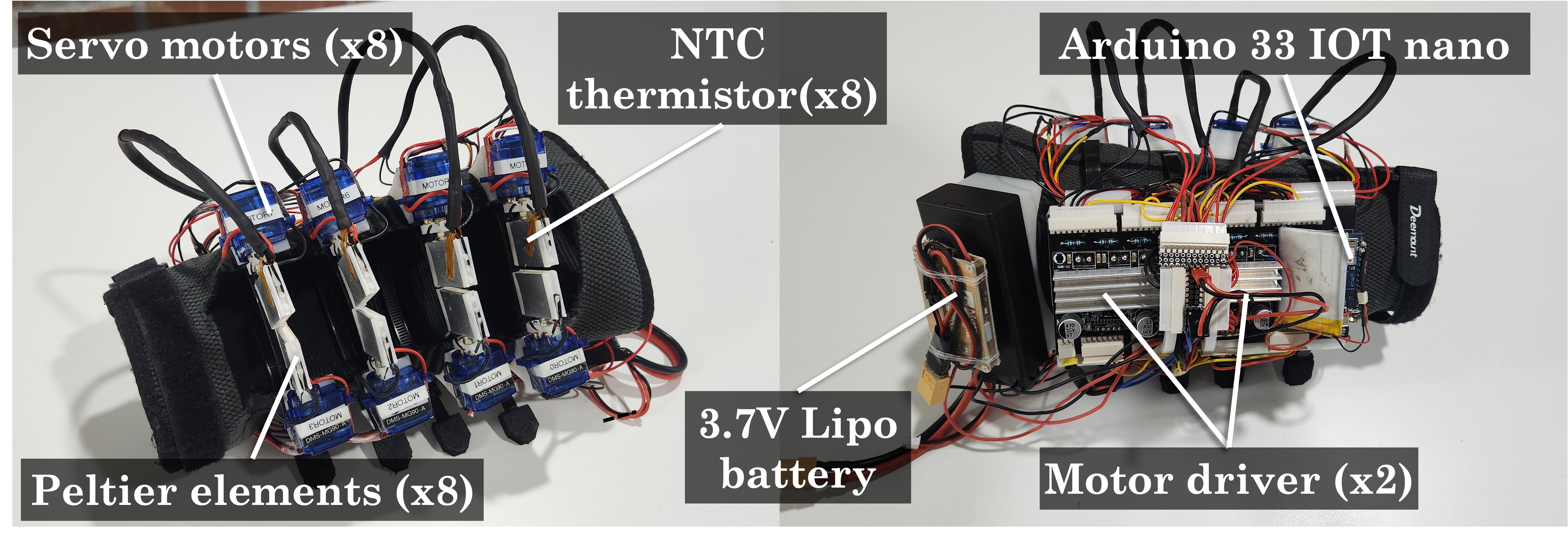}
      \caption{Hardware configuration of the our device. (Left) Front view showing servo motors, temperature sensors, and Peltier elements. (Right) Rear view showing the battery, motor driver, and microcontroller.}
      \label{fig:hardware_implementation}
      \vspace{-0.4cm}
   \end{figure}

\section{System Design and Implementation}
We integrated eight motor-driven Peltier elements, each capable of alternating warm and cold sensations by flipping the preheated or cooled side of the element (Fig \ref{fig:hardware_implementation}). The Peltier, mounted on a servo motor capable of 270-degree rotation, delivers the desired temperature swiftly, synchronized with the motor's reaction time. Elastic bands and rotation pivots were added to ensure consistent skin contact during direct Peltier and motor rotation. Each Peltier-motor pair is independently controlled by a motor driver, managed by an Arduino Nano 33 IOT, which also regulates the temperature.
   
   \begin{figure}[thpb]
      \centering
      \includegraphics[width=\linewidth]{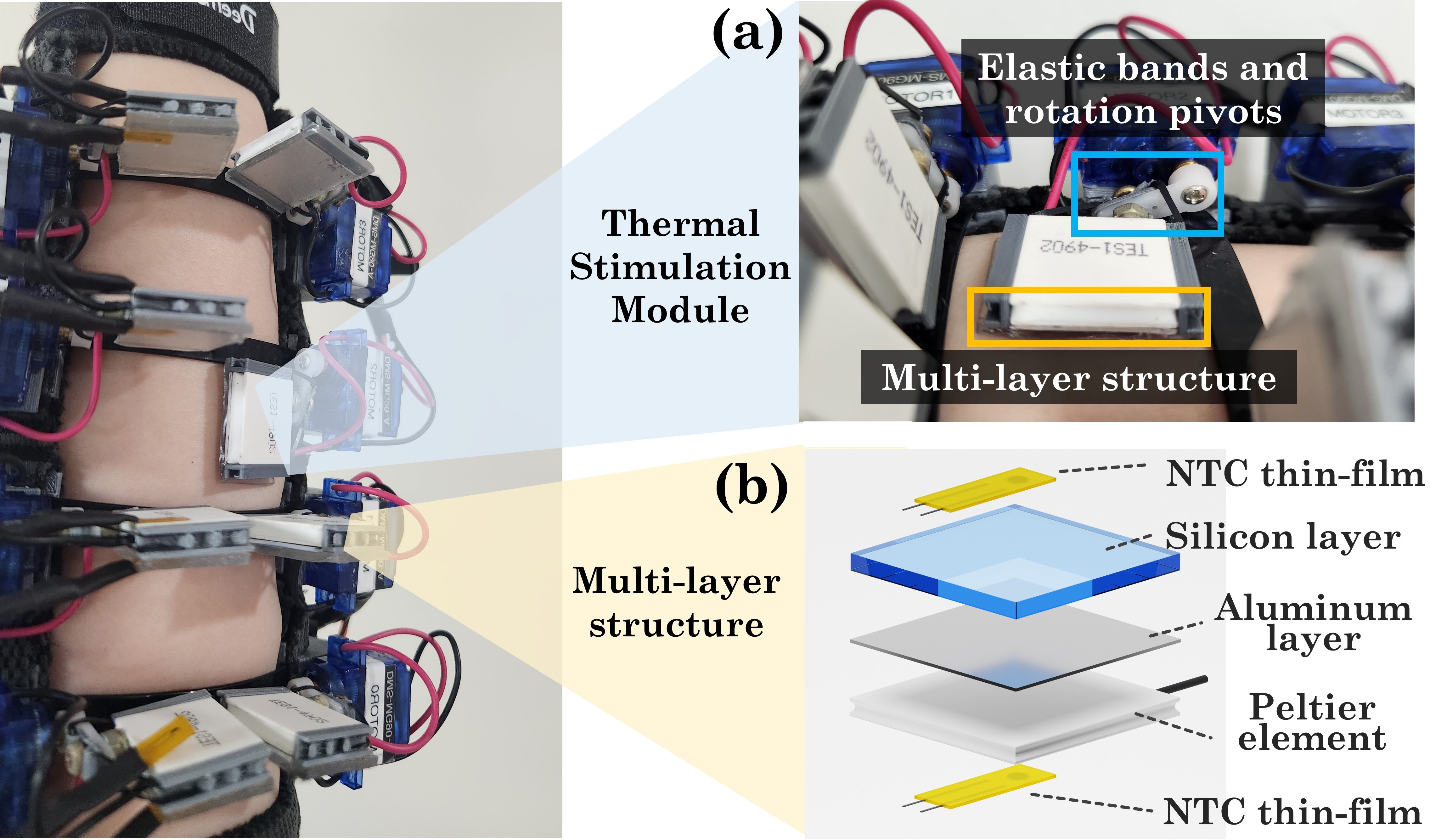}
      \caption{Detailed design of the dual-sided Peltier element. (a) Detailed design utilizing elastic bands and rotation pivots. (b) Multi-layer composition.}
      \label{fig:peltier_composition}
   \end{figure}

To incorporate both heating and cooling sides into XR feedback while ensuring user comfort and safety, we designed a multi-layer structure comprising a 1mm thick aluminum plate and a 3mm thick silicone layer (Fig \ref{fig:peltier_composition}). The aluminum layer, positioned on the warm side, minimizes excessive heat transfer, while the silicone acts as a thermal barrier, protecting the skin from direct contact. This configuration maintains the device's temperature, ensuring the safe delivery of thermal stimuli and preventing sudden temperature drops upon skin contact. Additionally, each group includes one Peltier element attached  with an NTC thin-film thermistor (MF5B 10K) on both sides for temperature control, specifically focused on maintaining warmth.

\section{Time-temperature Characteristics Analysis}
\begin{figure}[thpb]
      \centering
      \includegraphics[width=\linewidth]{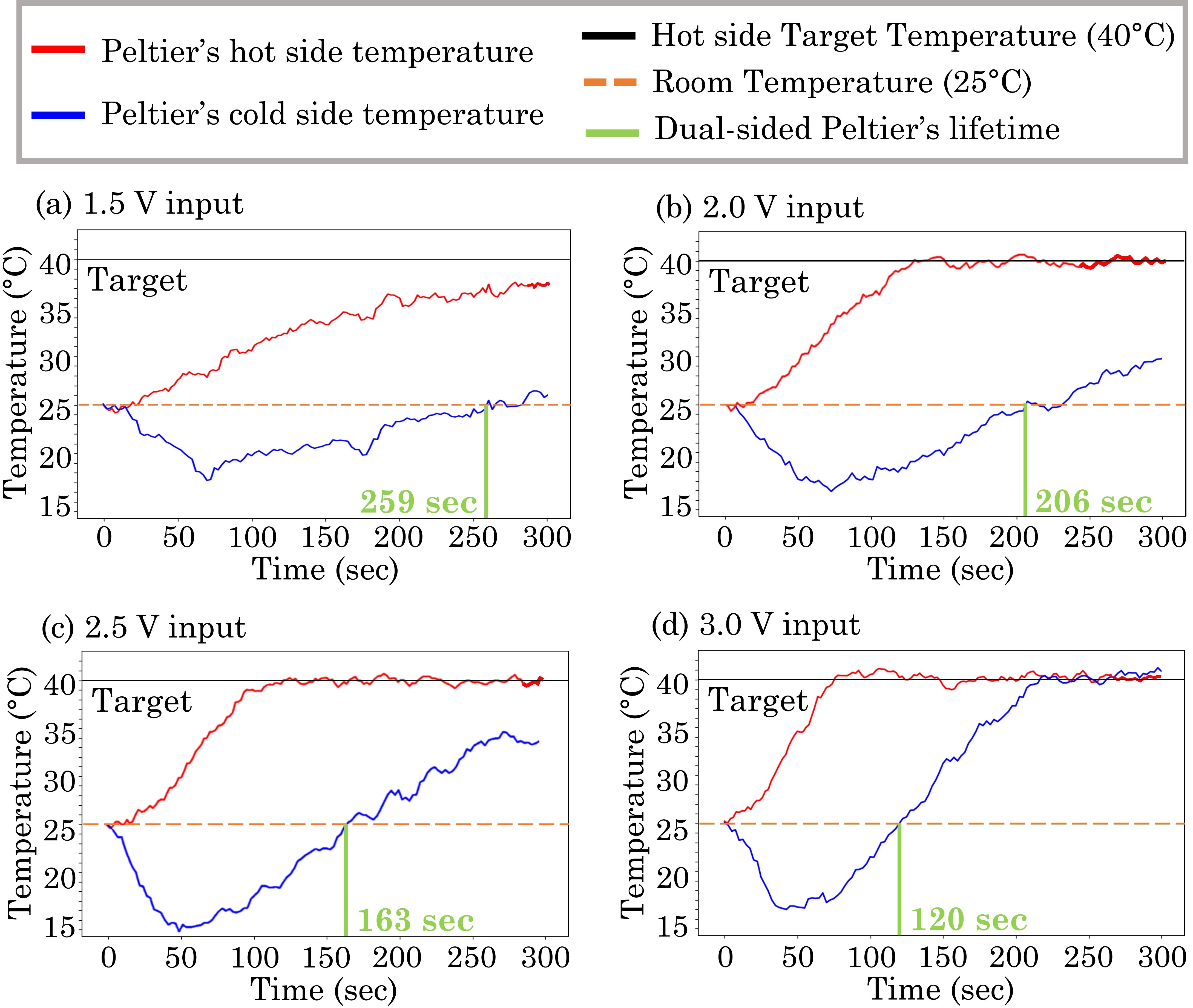}
      \caption{Thermal Parameter test for measuring the time-dependent temperature and lifetime of the dual-sided Peltier (input range 1.5-3.0V)}
      \label{fig:TemGraph}
   \end{figure}

We conducted a series of tests on TES1-4902 Peltier elements (20mm * 20mm, multi-layered), investigating their temperature changes over time with voltages ranging from 1V to 5V in 0.5V increments. Our aim was to achieve a warm side temperature of 40°C in a room maintained at 25°C. Each voltage level underwent three tests to assess crucial time-voltage dynamics (refer to Fig \ref{fig:TemGraph} for the results).

Initially, the warm side heated up while the cold side cooled down. However, heat transfer from warm to cold eventually caused the cold side to return to room temperature. We defined the "dual-sided Peltier element lifetime" as the duration until the cold side exceeded 25°C. At 1.5V, the warm side did not reach 40°C within the element's lifetime (\textit{M} = 259.2 s, \textit{SD} = 2.27 s). Starting from 2.0V, the target temperature was achieved before the end of the lifetime (\textit{M} = 206.3 s, \textit{SD} = 3.21 s), with decreasing lifetimes at higher voltages (2.5V: \textit{M} = 163.4 s, \textit{SD} = 1.25 s; 3.0V: \textit{M} = 120.1 s, \textit{SD} = 3.30 s). Therefore, we have set 2.0V as the optimal input voltage for achieving the target temperature with a warm sensation and for sustaining the maximum duration of a cool.

\section{DEMO APPLICATIONS}
\subsection{Dynamic Thermal Feedback in VR Gaming}
Haptic feedback, including vibration and pressure, has been shown to enhance immersion and presence in VR game scenarios \cite{kreimeier2019evaluation, hwang2024ergopulse, hwang2022reves}. However, thermal stimuli in response to virtual object interaction have been less explored due to the slow response time of Peltier-based systems. Our system enables rapid transitions between warm, neutral, and cold sensations, allowing for visually congruent thermal responses. For instance, content could include simulating showering with alternating warm and cold water (Fig \ref{fig:demoapplications} (a)).

\subsection{High-Resolution Thermal Information Display}
Our method's compact design, in contrast to techniques employing fluid in tubes \cite{gunther2020therminator} and chambers \cite{cai2020thermairglove}, enables scalability for matrix-wise information display, even in confined areas such as the forearm (Fig \ref{fig:demoapplications} (b)). Since each motor can be controlled and addressed independently, they can present information using both static and dynamic patterns. Thermal information inherently conveys a sense of atmosphere or context even prior to being encoded into specific data, thus offering potential as a channel for delivering warning information, which requires instantaneous stimulus delivery.

\subsection{Enhancing Telepresence with Sharing Body Temperature}
\begin{figure}[thpb]
      \centering
      \includegraphics[width=\linewidth]{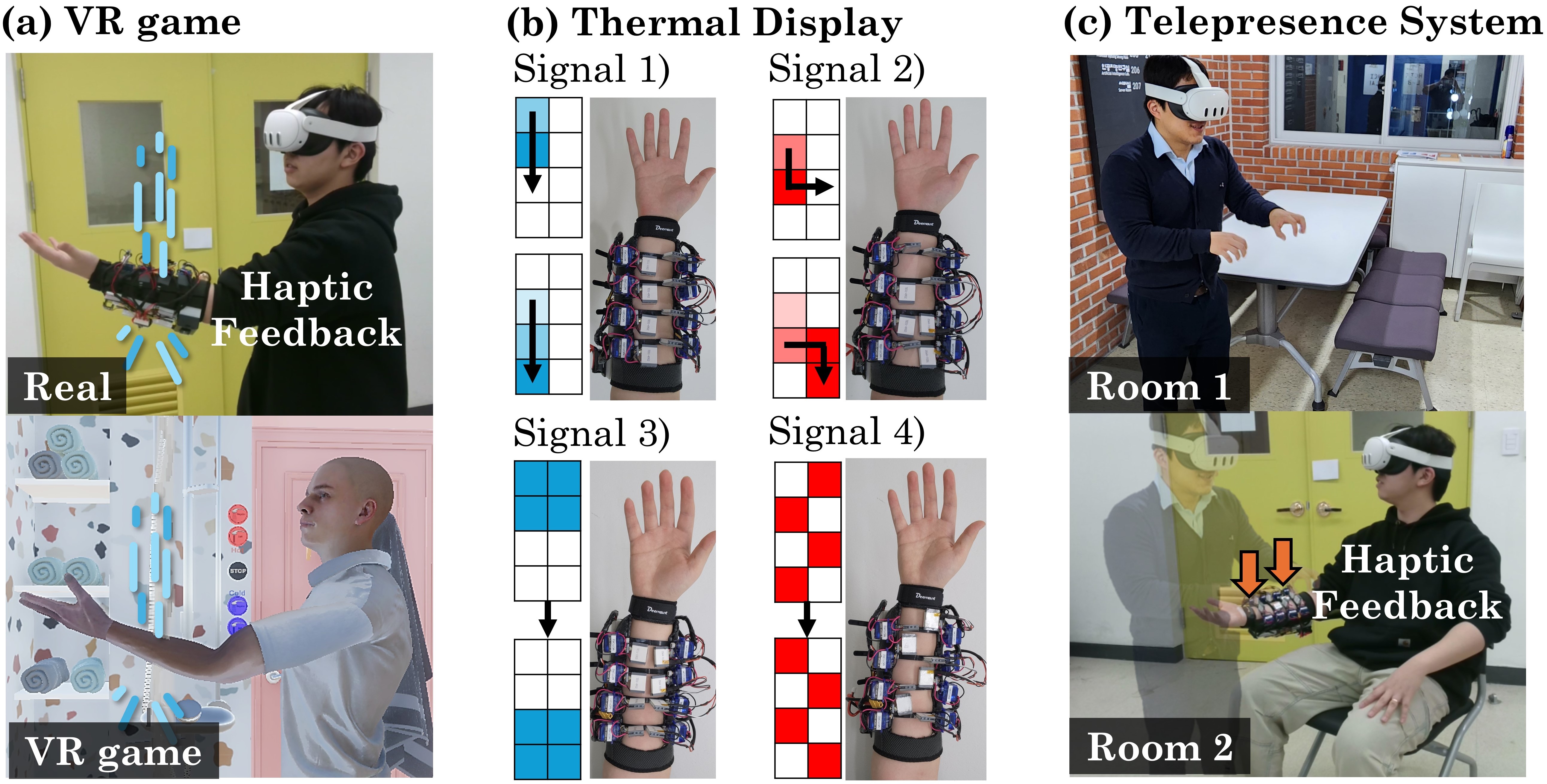}
      \caption{Demo applications where our system can be utilized: (a) a VR game with thermal feedback, (b) diverse thermal signals for thermal information display, (c) a telepresence system enabling to perceive touch and warmth.}
      \label{fig:demoapplications}
      \vspace{-0.1cm}
   \end{figure}

Telepresence applications utilize visual and auditory media to recreate physical presence and enable interactions over distances. Incorporating tactile and thermal feedback enhances the sense of presence and emotional exchange \cite{cascio2019social}. In this context, our system can deliver the warmth of the human body to remote users, fostering a heightened sense of interconnectedness during telepresence (Fig \ref{fig:demoapplications} (c)).

\section{FUTURE WORKS AND CONCLUSIONS}
We introduced motor-driven Peltier elements to deliver instant sensations of warmth and coolness in XR, overcoming the inherent slow response rate of Peltier elements. Our analysis of time-temperature characteristics demonstrated the feasibility of our multi-layer approach, utilizing both sides of the Peltier elements to generate warm and cold sensations efficiently. The immediate thermal transitions enabled by our system can benefit VR games requiring dynamic thermal feedback, high-resolution thermal displays for information delivery, and telepresence systems for natural and emotionally engaging communication with remote users. Moreover, our system is adaptable to other body parts such as feet and palms, enhancing interactions with virtual objects in XR. Given the heightened sensitivity of these body parts compared to forearms, future research could explore potential patterns a scaled-up version of our system could offer. Although our current prototype is somewhat bulky for wearable use, future developments could leverage soft actuators like SMA springs or electroactive polymers to achieve a lighter, wearable form.

\section{ACKNOWLEDGE}
This work was supported by the GIST-MIT Research Collaboration grant funded by the GIST in 2024.

\addtolength{\textheight}{-12cm}

\bibliographystyle{IEEEtran}
\bibliography{ref}

\end{document}